\documentclass[reprint,amsmath,amssymb,pra]{revtex4-2}
\usepackage[utf8]{inputenc}
\usepackage{graphicx}
\usepackage{dcolumn}
\usepackage{bm}
\usepackage{cases}
\usepackage{float}
\usepackage{mhchem}
\usepackage{ulem}
\usepackage{blindtext}
\usepackage{hyperref}
\hypersetup{
   pdfpagemode=UseNone, 
   pdfstartpage=1,
   pdfmenubar=true,
   pdftoolbar=true,
   colorlinks = true,
   linkcolor=blue,
   citecolor=blue,
   urlcolor=blue,
   bookmarksopen=false
 }

\begin{document}
\title{Temperature-induced miscibility of impurities in trapped Bose gases}
\author{G. Pascual$^1$, T. Wasak$^2$, A. Negretti$^3$, G. E. Astrakharchik$^1$ and J. Boronat$^1$}
\affiliation{
$^1$ Departament de F\'{i}sica, Universitat Polit\`ecnica de Catalunya, Campus Nord B4-B5, E-08034, Barcelona, Spain \\
$^2$ Institute of Physics, Faculty of Physics, Astronomy and Informatics, Nicolaus Copernicus University in Toru\'n, Grudzi\c{a}dzka 5, 87-100 Toru\'n, Poland \\
$^3$ Zentrum für Optische Quantentechnologien, Fachbereich Physik, Luruper Chaussee 149, D-22761, Hamburg, Germany
}

\date{\today}
\begin{abstract}
We study the thermal properties of impurities embedded in a repulsive Bose gas
under a harmonic trapping potential. 
In order to obtain exact structural properties in this inhomogeneous many-body system, we resort to the path-integral Monte Carlo method. 
We find that, at low temperatures, a single impurity is expelled to the edges of the bath cloud if the impurity-boson coupling constant is larger than the boson-boson one. 
However, when the temperature is increased, but still in the Bose-condensed phase, 
the impurity occupies the center of the trap and, thus, the system becomes 
miscible. This thermal-induced miscibility crossover is also observed for a finite 
concentration of impurities in this inhomogeneous system. We find that the transition temperature for 
miscibility depends on the impurity-boson interaction and 
we indicate a different \textit{nondestructive} method to measure the 
temperature of a system based on the studied phenomenon.
\end{abstract}

\maketitle

\noindent \textit{Introduction.} 
The Bose polaron has been extensively studied both at zero temperature \cite{Rath2013,Li2014,Grusdt2015,Ardila2015,Volosniev2015,Levinsen2015,Shchadilova2016,Jorgensen2016,Hu2016,Ardila2016,Sun2017,Yoshida2018,Loon2018,Mistakidis2019,Ichmoukhamedov2019,Drescher2020,Levinsen2021,Massignan2021,Isaule2021,Christianen2022,Christianen2022PRA,Skou2022} as well as at finite temperatures \cite{Tempere2009,Guenther2018,Pastukhov2018,Field2020,Zoe2020,Pascual2021}.
Such a system presents a significant interest not only due to fundamental questions about quasi-particle formation, but also the mobility of the impurity 
inside the Bose-Einstein condensate (BEC), as was recently shown, can serve as a \textit{nondestructive} 
indicator of the temperature of the whole system~\cite{Mehboudi2019}. However, the question of how a single impurity behaves in an inhomogeneous BEC at finite temperature was not addressed.

In homogeneous repulsive Bose-Bose mixtures, the mean-field approximation provides a robust criterion to distinguish miscible and immiscible phases at zero temperature. In particular, if $g_{BB}$, $g_{II}$, and $g_{BI}$ denote the intra- and inter-species coupling constants respectively, the two components of the mixture are phase separated when $g_{BI}>\sqrt{g_{BB}g_{II}}$ and they are mixed when $g_{BI}<\sqrt{g_{BB}g_{II}}$~\cite{Ao1998}. Quantum Monte Carlo (QMC) techniques have verified this criterion and, moreover, they have extended the analysis to finite temperatures revealing that Popov theory fails in describing repulsive Bose mixtures at finite polarization~\cite{Ota2019,Ota2020,Spada2023,Pascual2023}. Nevertheless, an inhomogeneous thermal mixture has only been studied, using QMC techniques, in~\cite{Dzelalija2020} with the same number of particles per species.

The behavior of an impurity in an inhomogeneous gas at finite temperature and the effect of the number of impurities in the miscibility of the system are topics that have not been addressed in the field yet. QMC techniques can shed some light on it and provide novel methods for thermometry.

\begin{figure}[!]
\centering
\includegraphics[width=7cm,height=7cm]{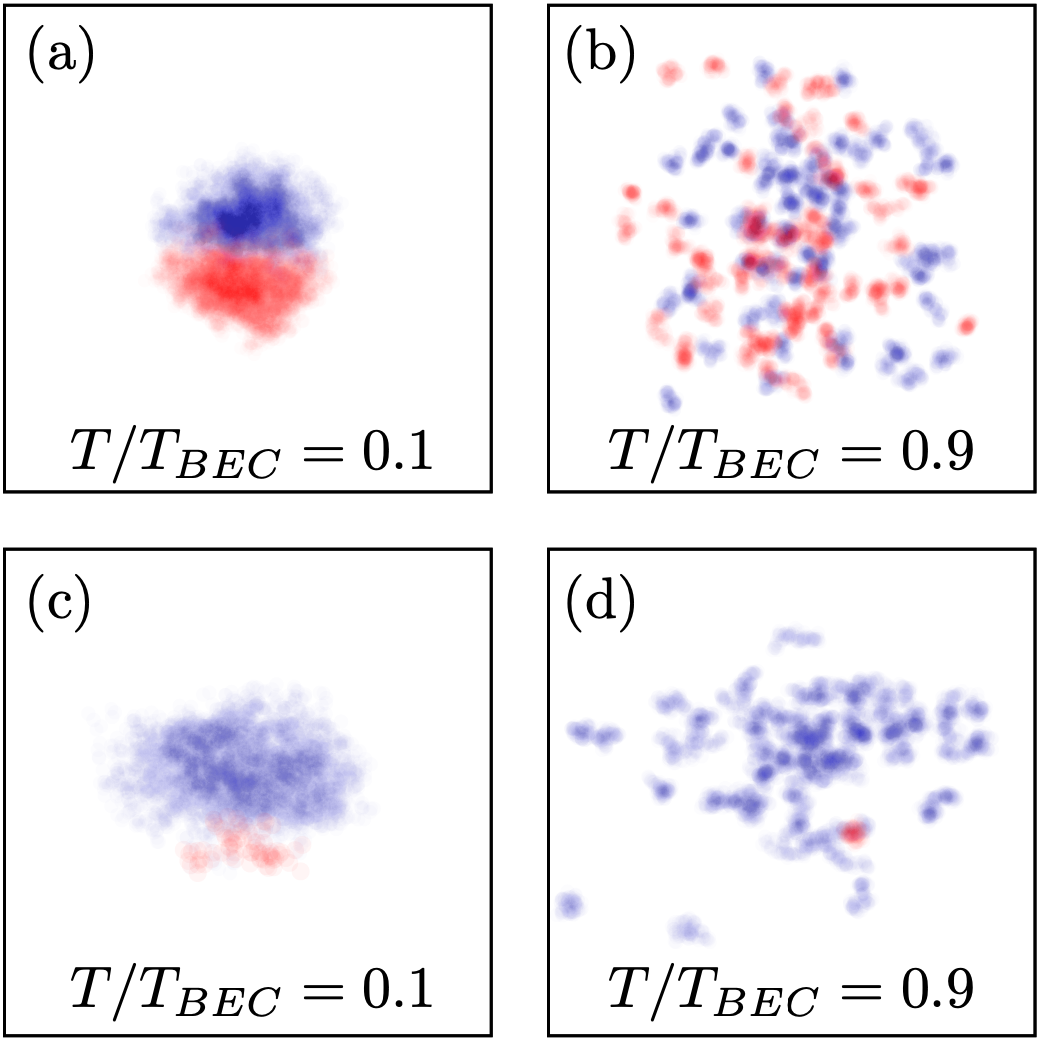}
\caption{Snapshots of PIMC positions of the {\it beads} of particles (blue) and impurities (red) for different characteristic parameters. 
Panels~(a) and (b) correspond to a binary mixture with the number of particles equal to the number of impurities, $N_\text{I} = N$,  $g_{BI}/g = 3.5$ and $g_{BB}=g_{II}=g$. 
Panels~(c) and (d) show a single impurity with $g_{BI}/g = 4.5$ in a bath.
The systems are immiscible at $T/T_{\text{BEC}} = 0.1$ and are miscible at $T/T_{\text{BEC}} = 0.9$, being $T_{\text{BEC}}$ defined in Eq.~(\ref{eq:T_bec}).
}
\label{fig:position}
\end{figure}

In the present work, our main goal is to study thermal properties of impurities immersed into an inhomogeneous Bose gas.  We use the Path Integral Monte Carlo method (PIMC) which provides an exact technique to find structural properties such as the density profile in an exact way, within controllable statistical errors. We analyze the system for different interaction strengths between the impurity and the rest of the particles, and we observe that above a certain interaction $g_{BI}$ the impurity remains outside the Bose gas (see Fig.~\ref{fig:position}). Finally, we observe the existence of a mixing temperature above which the impurity penetrates the center of the gas.

\noindent \textit{Model.}
The Hamiltonian of the $(N+N_I)$-particle system in a harmonic trap in three dimensions is given by
\begin{eqnarray}\label{eq:Hamiltonian}
&H& = -\frac{\hbar^2}{2m}\sum_{i=1}^{N}\nabla^2_i - 
\frac{\hbar^2}{2m}\sum_{I=1}^{N_I}\nabla^2_{I} + \sum_{i<j}^{N} V_{BB}(r_{ij}) 
\\
&+& \!\sum_{i,I}^{N,N_I}\!\!V_{BI}(r_{iI})\!+\!\sum_{I<J}^{N_I}\!\!V_{II}(r_{IJ})\!+\!\!\sum_{n=1}^{N} V_{\text{ext}}(r_{i})\!+\!\sum_{I=1}^{N_I}\!V_{\text{ext}}(r_{I}),
\nonumber
\end{eqnarray}
where lower (upper) case indices denote particles (impurities) with positions $r_{\alpha}$ and distances $r_{\alpha\beta} = |r_{\alpha} - r_{\beta}|$ between them, with $\{\alpha,\beta\} = \{i,I\}$.
We consider an experimentally relevant case where all the particles have the same mass $m$. We model the repulsive boson-boson (impurity-impurity) interaction potential by $V_{BB}(r)= V_{II}(r) = V_0/r^{12}$ and boson-impurity by $V_{BI}=V_0^{BI}/r^{12}$ with amplitudes $V_0$ and $V_0^{BI}$ chosen to reproduce~\cite{Pilati2006,Landau1977} the desired values of the $s$-wave scattering length $a$ and $a_{BI}$, correspondingly. The coupling constants, which appear in the mean-field theory, $g=g_{BB}=g_{II}$ and $g_{BI}$ are related to $a$ and $a_{BI}$ within the first Born approximation as $g=4\pi\hbar^2a/m$, and similarly for $g_{BI}$. 
The external confinement is taken in the form of a harmonic trap with frequency $\omega$, i.e.,  $V_{\text{ext}}(r) = m\omega^2 r^2/2$,
which defines a characteristic oscillator length scale set by $a_{\text{ho}} = \sqrt{\hbar/(m\omega)}$. 
For sufficiently low values of the gas parameter, $na^3$, the specific shape of the interaction potential is no longer important and the description in terms of the $s$-wave scattering length becomes universal~\cite{Giorgini1999}.
To ensure that the simulations are performed in this universal regime, we ascertain that the density $n_0$ at the center of the trap in our simulations is sufficiently low, $n_0a^3<10^{-5}$. 
This is achieved by setting $a_{\text{ho}}/a = 15$ while using around one 
hundred particles.
The temperature scale is set by the degeneracy temperature of an ideal 
BEC in a harmonic trap,
which depends solely on the number of particles $N$ and the trap frequency $\omega$, i.e.,
\begin{equation}\label{eq:T_bec}
k_B T_{\text{BEC}} = \hbar \omega \left(\frac{N}{\zeta(3)}\right)^{1/3},
\end{equation}
where $\zeta(x)$ is the Riemann zeta function.

\noindent \textit{Method.} 
We perform PIMC simulations of $N$ bosons and $N_I$ impurities in a 3D harmonic trap, see Eq.~\eqref{eq:Hamiltonian}. In this method, the kinetic and the potential parts of the temperature density matrix are decoupled and a {\it trotterization} algorithm is applied~\cite{Ceperley1995}. In this scheme, the exponential of the Hamiltonian is divided into small divisions (called 
{\it beads}) in such a way that the thermal density matrix can be approximated into decoupled terms by using fourth-order expansions such as the Chin action~\cite{Chin2002,Sakkos2009}. This technique has been used in other works exploring dilute mixtures at finite temperature in which structural properties of the system have been computed accurately~\cite{Pascual2021,Pascual2023}. The indistinguishability of the bosonic particles is imposed by sampling permutations using the worm algorithm~\cite{Boninsegni2006}. In this algorithm, particles, represented as polymers, where each \textit{subparticle} corresponds to a different \textit{bead} (i.e., the particle at a particular imaginary time), can be cut, bound and swapped to other polymers making the overall chain a cluster of indistinguishable particles. We note that we have verified that doubling the number of particles does not lead to any significant change in the observed phenomena.

\begin{figure}[!]
\centering
\includegraphics[width=8.6cm,height=7.8cm]{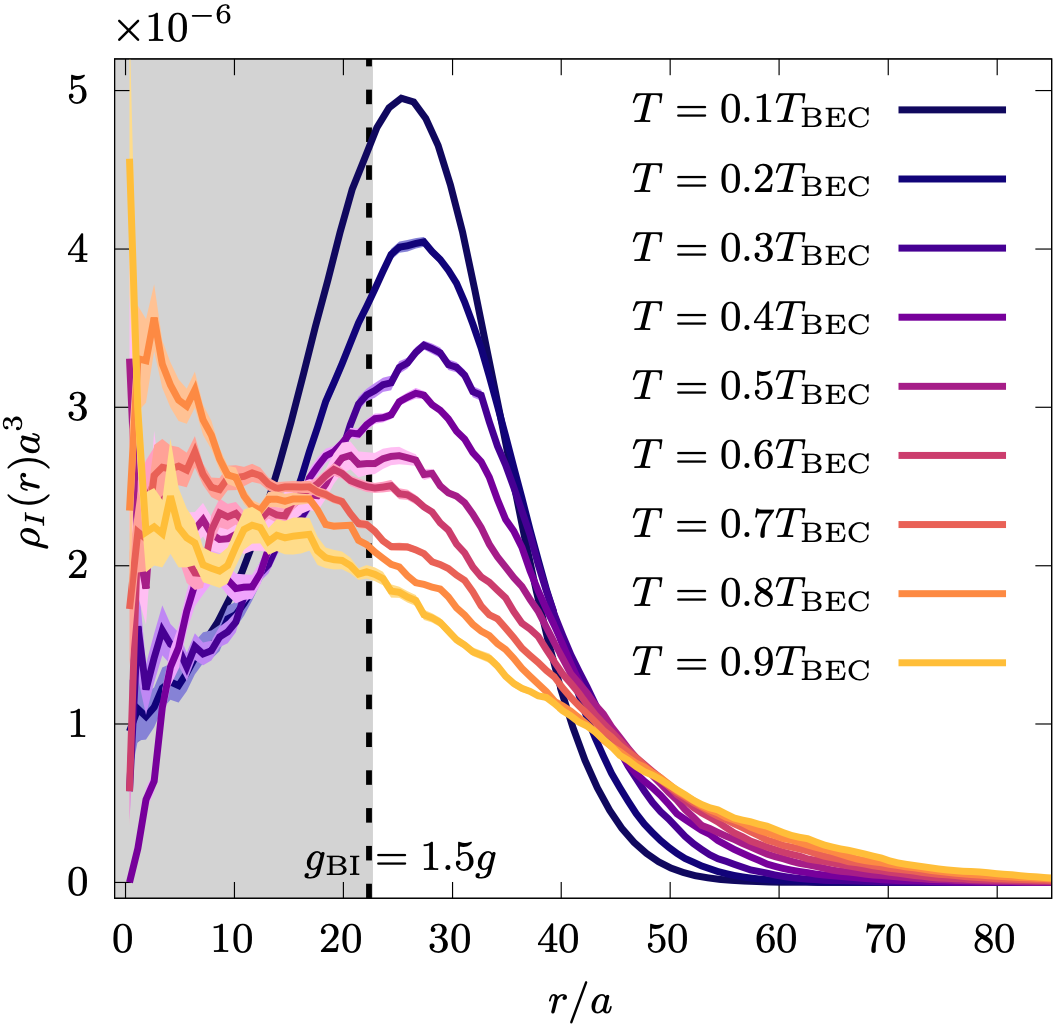}
\caption{Single impurity density profile in a bath containing $N=64$ bosons with $g_{BI}/g = 1.5$ and $a_{\text{ho}}/a = 15$ is depicted using various colors, each corresponding to a different temperature. The density profile is normalized to the number of impurities ($N_I=1$ in this case). The dashed vertical line shows the most probable impurity position as approximated by Eq.~\eqref{eq:radius_imp}. The gray area shows where half of the particles of the bath are located at $T=0.1T_{\text{BEC}}$ (in Appendix \ref{sec:App_DP_Bath} we show the density profiles). Notice that the impurity starts penetrating the center of the trap as the temperature is increased.
}
\label{fig:density_main}
\end{figure}

\begin{figure*}[!]
\centering
\includegraphics[width=16cm,height=6cm]{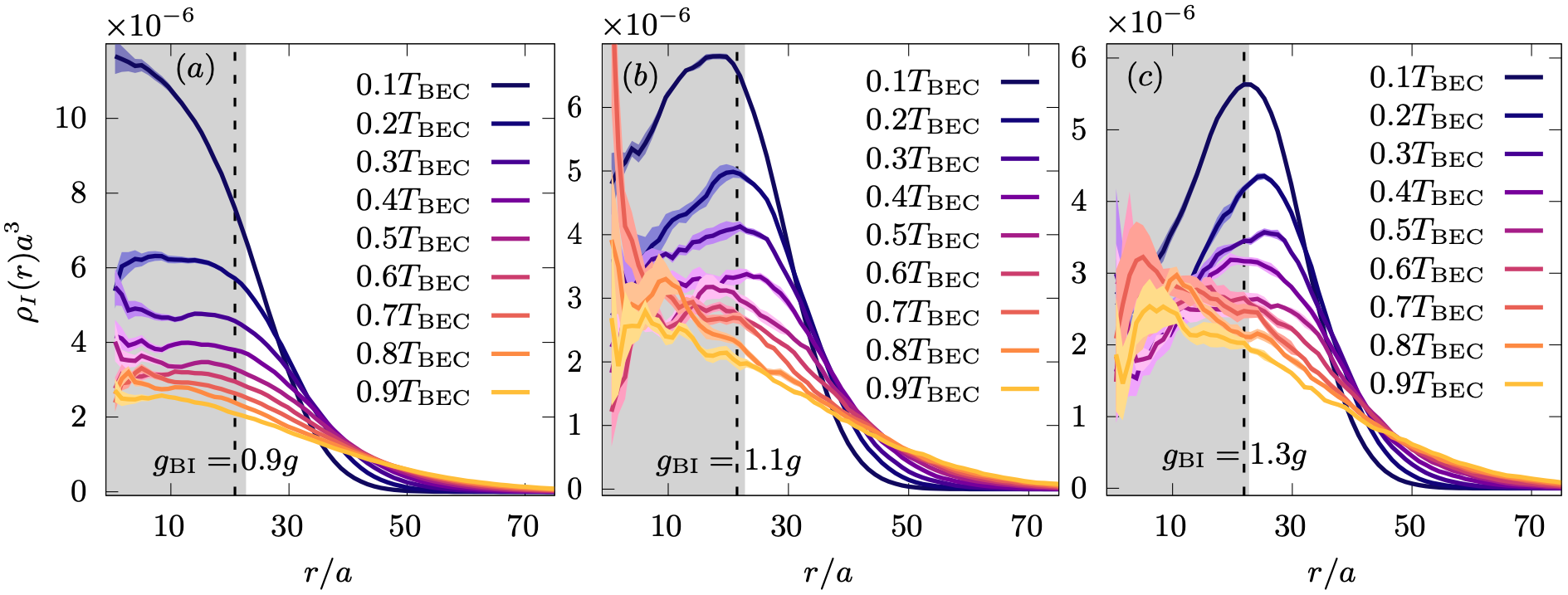}
\caption{Single-impurity density profile, similar to Fig.~\ref{fig:density_main} but for various values of $g_{BI}/g$, including values both larger and smaller than 1.
}
\label{fig:diff_a12}
\end{figure*}

\begin{figure*}[!]
\centering
\includegraphics[width=16cm,height=6cm]{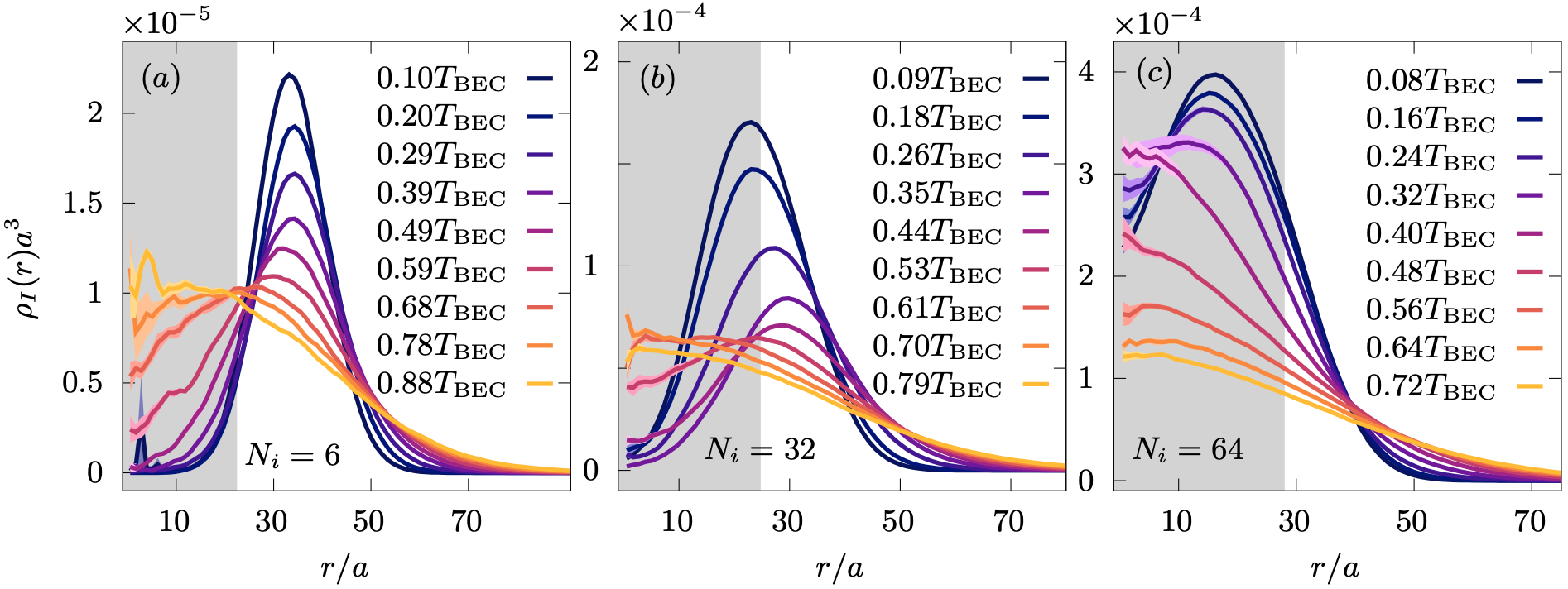}
\caption{Multiple-impurity density profile in a bath containing $64$ bosons with $g_{BI}/g = 3.5$ and $a_{\text{ho}}/a = 15$. The density profiles are normalized to the number of impurities ($N_I = 6; 32; 64$), in increasing order from left to right. The gray area shows where half of the particles of the bath are located at $T=0.1T_{\text{BEC}}$.}
\label{fig:more_imp}
\end{figure*}

\noindent \textit{Results.} 
In order to study the effects of the temperature, we focus on experimentally observable structural properties, such as the density profile of the impurity. The density profiles shown in the following figures have been all computed using the PIMC method explained above. In Fig.~\ref{fig:density_main}, we report for a system with a single impurity ($N_I = 1$) various density profiles
obtained by changing the temperature while keeping the ratio of the coupling 
constants fixed to $g_{BI}/g = 1.5$. Remarkably, at low temperatures, the impurity 
remains mostly outside of the central region of the trap in which bath density 
is maximal. This effect emerges from a delicate balance between the repulsive 
interactions and potential energy of the external harmonic trapping, since by 
expelling the impurity, its energy associated with the external potential is 
increased, while the energy coming from the interaction with bath particles 
is decreased. The contribution of the impurity to the potential energy can be roughly 
estimated by assuming that the boson-impurity interaction is weak and the 
impurity profile $\rho_I({\bf r})$ does not modify the density profile of the 
bath $\rho_B({\bf r})$
\begin{eqnarray}
E_{I} &=& \int d{\bf r} \int d{\bf r}_I \rho_{\text{B}}({\bf r}) \rho_{\text{I}}({\bf r}_I) V_{BI}(|\bf r-\bf r_I|) \nonumber \\ 
&+& \int d{\bf r} \rho_{\text{I}}({\bf r}) V_{ext}({\bf r}) \ ,
\label{aproxrho}
\end{eqnarray}
where $\rho_B({\bf r})$ can be taken as the ideal Bose gas density at $T=0$, i.e., $\rho^{\text{IG}}_{\text{B}}(r) = N/(\pi^{3/2}a_{\text{ho}}^3)\;\exp(-r^2/a_{\text{ho}}^2)$. Equation~\eqref{aproxrho} can be further simplified by assuming that the boson-impurity interaction is a contact pseudopotential, $V_{\text{BI}}({\bf r})\to g_{\text{BI}}\delta({\bf r})$ and the impurity is located at a certain distance from the center, $\rho_\text{I}({\bf r}) \propto \delta (x - r_I)\delta(y)\delta(z)$, where we arbitrarily choose the direction of the impurity as the $x-$axis. Within these approximations, one gets
\begin{equation}
\label{eq:E:impurity}
E_\text{I} = g_\text{BI}\;\rho_\text{B}^\text{IG}(r_\text{I}) + V_\text{ext} (r_\text{I}).
\end{equation}

The most probable position of the impurity, $r_\text{I}$, is then defined by 
minimization of its potential energy~(\ref{eq:E:impurity}), 

\begin{equation}\label{eq:radius_imp}
\frac{r_I}{a_{\text{ho}}} = \sqrt{\ln \left( \frac{8}{\pi^{1/2}}\frac{N a_{BI}}{a_{\text{ho}}}\right)}.
\end{equation}

This prediction for the most probable position of the impurity is shown in Fig.~\ref{fig:density_main} with a vertical dashed line. A reasonably good agreement is found in the cases when the impurity is expelled from the center, which happens for low $T$. For strong boson-impurity interactions (i.e. $g_{BI}/g \gg 1$), Eq. (\ref{eq:radius_imp}) loses accuracy because it is derived by assuming an absence of correlations and, moreover, the impurity starts to significantly modify the density profile of the bath.

Another limiting factor to our approximation is that in Eq.~\eqref{eq:radius_imp} we assumed a non-interacting bath. However, for a significantly larger number of particles, such that $(N a/a_{\text{ho}})^{1/5}\gg 1$, the local density approximation (LDA) is expected to be applicable and the zero-temperature density profile of the bath would instead take a form of an inverted parabola~\cite{Pitaevskii2016}.

As the temperature of the system is increased, we find that at a mixing temperature ($T=T_m$) 
the impurity fully penetrates the center of the trap. We define $T_m$ as the temperature at which the peak in the density profile of the impurity moves from a finite value to the center of the trap. The mixing temperature $T_m$ depends on the coupling constant $g_{BI}$; for example, we find $T_m/T_{\text{BEC}} = 0.6 \pm 0.1$ for $g_{BI}/g = 1.5$, and $T_m/T_{\text{BEC}} = 0.8 \pm 0.1$ for $g_{BI}/g = 4.5$. We emphasize that in all our calculations  $T_m$ is lower than the degeneracy temperature $T_{\text{BEC}}$, which means that this phenomenon appears in the condensed phase.

Since the mean-field criterion for a miscible to immiscible transition, expressed as $g_{BI} = \sqrt{g_{BB}g_{II}}$, is not applicable to a single impurity due to the absence of a defined value of $g_{II}$, we investigate miscibility as a function of the ratio $g_{BI}$ and $g = g_{BB}$.
To this end, in Fig.~\ref{fig:diff_a12} we plot the density profiles of the impurity for various $T$ and $g_{BI}/g$. 
We find that when $g_{BI}/g < 1$ the impurity remains miscible with the rest of the particles at the center of the trap while, for $g_{BI}/g > 1$, we observe the same immiscible behavior as in Fig.~\ref{fig:density_main}, i.e. at low $T$ the impurity is repelled from the central region of the trap and it penetrates the center when the temperature increases. We note that the estimated position of the impurity with Eq.~(\ref{eq:radius_imp}) works well in the immiscible phase while it fails in the miscible one.

\begin{figure}[!]
	\centering
	\includegraphics[width=8.6cm,height=11cm]{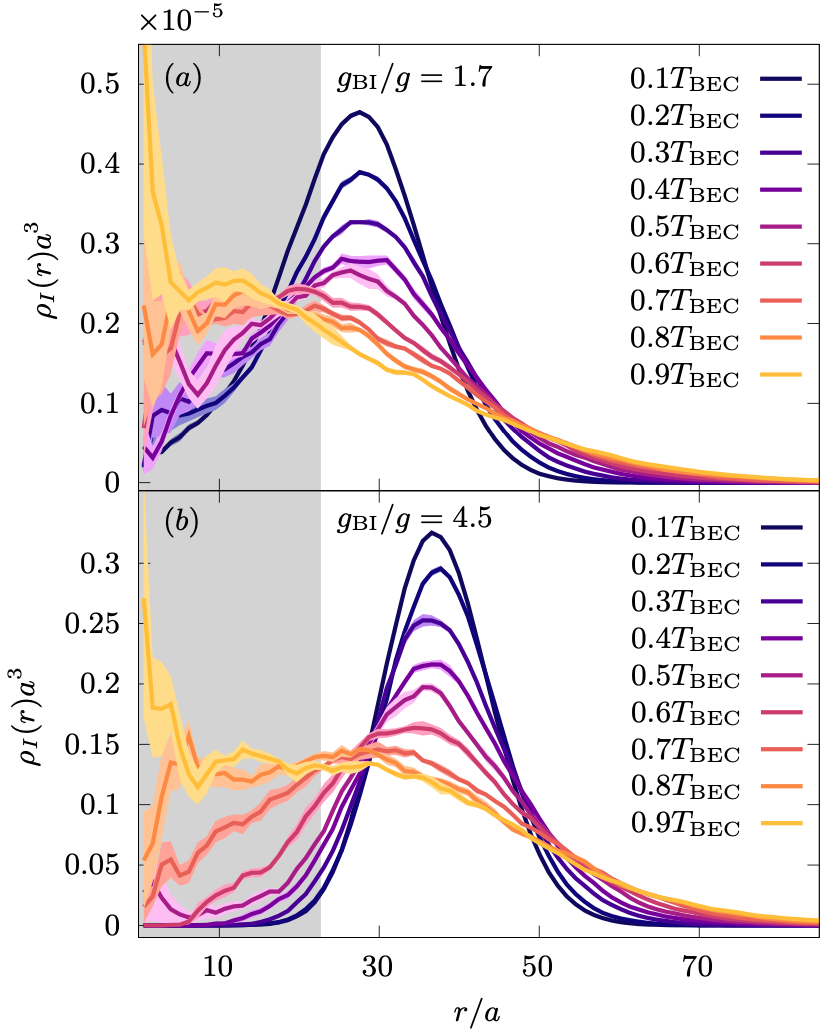}
	\caption{Density profiles of the impurity interacting repulsively with a bath of $64$ bosons at different $g_{BI}/g$ and with $a_{\text{ho}}/a = 15$. The density profile is normalized to the number of impurities ($N_I=1$ in this case). The gray area shows where half of the particles of the bath are located at $T=0.1T_{\text{BEC}}$.}
    \label{fig:diff_a12_higher}
\end{figure}

To explore the influence of multiple impurities on temperature-induced miscibility, we study $N_I > 1$ impurities interacting with each other with the same inter-atomic potential as the particles in the bath, i.e., $V_{II}(r) = V_{BB}(r)$. In Appendix \ref{sec:App_Finite_Size_Effets}, we study the effect of $N$ on the system.
In Fig.~\ref{fig:more_imp}, we show characteristic density profiles of the impurities for various number of impurities $N_I$, focusing on the case of strong atom-impurity interactions ($g_{BI}/g = 3.5$). 
At low temperatures the immiscible phase is formed. 
For small impurity concentration, the impurities are expelled outside creating a spherically symmetric shell and completely voiding the central region at $r=0$ (see Fig.~\ref{fig:more_imp}a), similarly to the single-impurity case. 
As the impurity concentration is increased, the outer shell broadens due to impurity-impurity interactions, as can be seen by comparing Fig.~\ref{fig:more_imp}a and Fig.~\ref{fig:more_imp}b, while the spherical symmetry is still preserved. 
Instead, in the balanced case, $N_I = N$, a phase with a different symmetry is realized~\cite{Cikojevic2018,Dzelalija2020}, composed of two blobs of pure phase, each occupying one half of the trap.
The interface surface between impurities and particles is minimized as it corresponds to a flat circle rather than to a sphere.
A typical snapshot of the two-blob phase is shown in Fig.~\ref{fig:position}a and it has $r=0$ region fully accessible for the impurities as manifested by large values of $\rho_I(r=0)$ (Fig.~\ref{fig:more_imp}c).
As the temperature is increased, the system changes its symmetry and undergoes a crossover to a miscible phase.

Finally, in Fig.~\ref{fig:diff_a12_higher} we study strongly repulsive interactions between the impurity and the rest of the particles. For large $g_{BI}/g$ ratios, the strong boson-impurity repulsion causes the impurity to be expelled to a larger radius, as can be expected from Eq.~(\ref{eq:radius_imp}). At the same time, more energy is needed to overcome the interaction with the bath (i.e. to occupy the center of the trap) and, as a consequence, the mixing temperature $T_m$ grows. This result indicates that $T_m$ has a significant dependence on~$g_{BI}$.

\noindent \textit{Conclusions and Discussion.} 
In this work, we address the polaron problem by performing PIMC simulations of impurities immersed in a bosonic gas within a harmonic trap. We compute the density profile of the impurity at finite temperatures.
We find that even though the mean-field criterion for homogeneous mixtures~\cite{Ao1998} is not defined in the single-impurity case, we observe miscible and immiscible regimes. 
At low temperatures, the impurity fully penetrates the central region of the bath cloud for $g_{BI}/g < 1$, while the bosonic bath expels the impurity to the outer shell for $g_{BI}/g > 1$.
Notably, this result goes beyond the quasi-particle picture where the impurity is surrounded and dressed by excitations of the bath. 
The observed phenomenon recalls what happens in droplets of \ce{^{4}He} when impurities of \ce{^{3}He} are added to the system~\cite{Krot1,Krot2}. 
These impurities are also expelled to the surface of the droplet creating what is known as the Andreev state~\cite{Andreev1969,Lekner1970}. 
Furthermore, we have found a closed expression that approximates the most probable distance of the impurity from the center. 
This prediction holds well for the immiscible phase ($g_{BI}/g > 1$) as long as the repulsion is not very strong.
Instead, for strong impurity-atom interactions and large impurity concentration the system loses spherical symmetry and splits into two blobs.

As the temperature of the system is increased, we observe a critical temperature $T_m$ above which the mixing occurs and the impurity fully penetrates the center of the trap. 
We discover a noticeable dependence of $T_m$ on the coupling constant $g_{BI}$. 
The more repulsive is the impurity, the higher is the mixing temperature $T_m$. This phenomenon can be used in experiments to estimate the temperature of the system without disturbing it. Notice that, as it has been shown in our work, the insertion of the impurity to the system barely modifies the properties of the bath and, thus, this method is \textit{nondestructive}. The relation between impurity properties $T_m$ can be obtained using QMC techniques and, if it is necessary, one can extrapolate the results to systems with a larger number of particles. 

The phenomenon identified in this study as well as its relation to thermometry can be experimentally 
observed since, with the current technologies, the properties of the impurities, e.g., its mean position, 
can be reliably measured. A possible challenge is to introduce only a single impurity to the system but, as we have shown, even with the presence of a few more impurities, the temperature-induced miscibility still persists making this approach a promising technique for measurements of the temperature in trapped ultracold gases.

\noindent\textit{Acknowledgments.}
This work has been supported by the Spanish Ministry of University under the 
grant FPU No.~FPU20/00013, the Spanish Ministry of Economics, Industry and 
Competitiveness under grant No.~PID2020-113565GB-C21, and
AGAUR-Generalitat de Catalunya Grant No. 2021-SGR-01411.
This research is part of the project No. 2021/43/P/ST2/02911 co-funded by the National Science Centre and the European Union Framework Programme for Research and Innovation Horizon 2020 under the Marie Skłodowska-Curie grant agreement No. 945339.
For the purpose of Open Access, the author has applied a CC-BY public copyright licence
to any Author Accepted Manuscript (AAM) version arising from this submission.

\textit{Data availability}: The data presented in this article is available from~\cite{SRFRWA_2024}.

\appendix

\section{Density profiles of the bath}\label{sec:App_DP_Bath}
\begin{figure}[h]
	\centering
	\includegraphics[width=8.6cm,height=7.8cm]{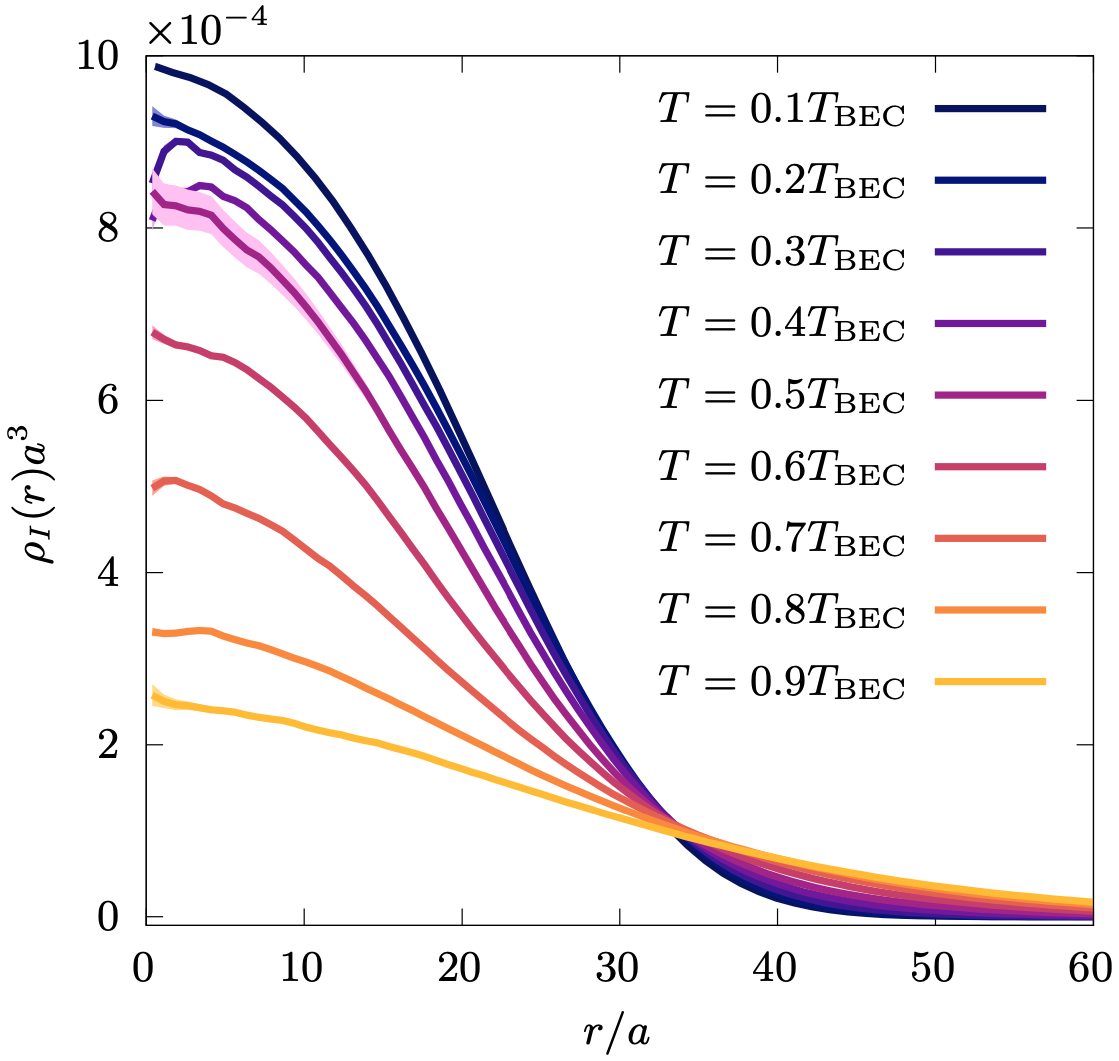}
	\caption{Density profiles of the particles in the bath at different temperatures with $g_{BI}/g = 1.5$ and $a_{\text{ho}}/a = 15$.}
    \label{fig:density_bath}
\end{figure}

In Fig.~\ref{fig:density_bath}, we show the density profiles of the particles in the bath at different temperatures. They are normalized to the number of particles ($N=64$). Notice that the peak of the density profile remains always at the center of the trap.

\section{Finite size effects} \label{sec:App_Finite_Size_Effets}
\begin{figure}[h]
	\centering
    \includegraphics[width=8.6cm,height=11cm]{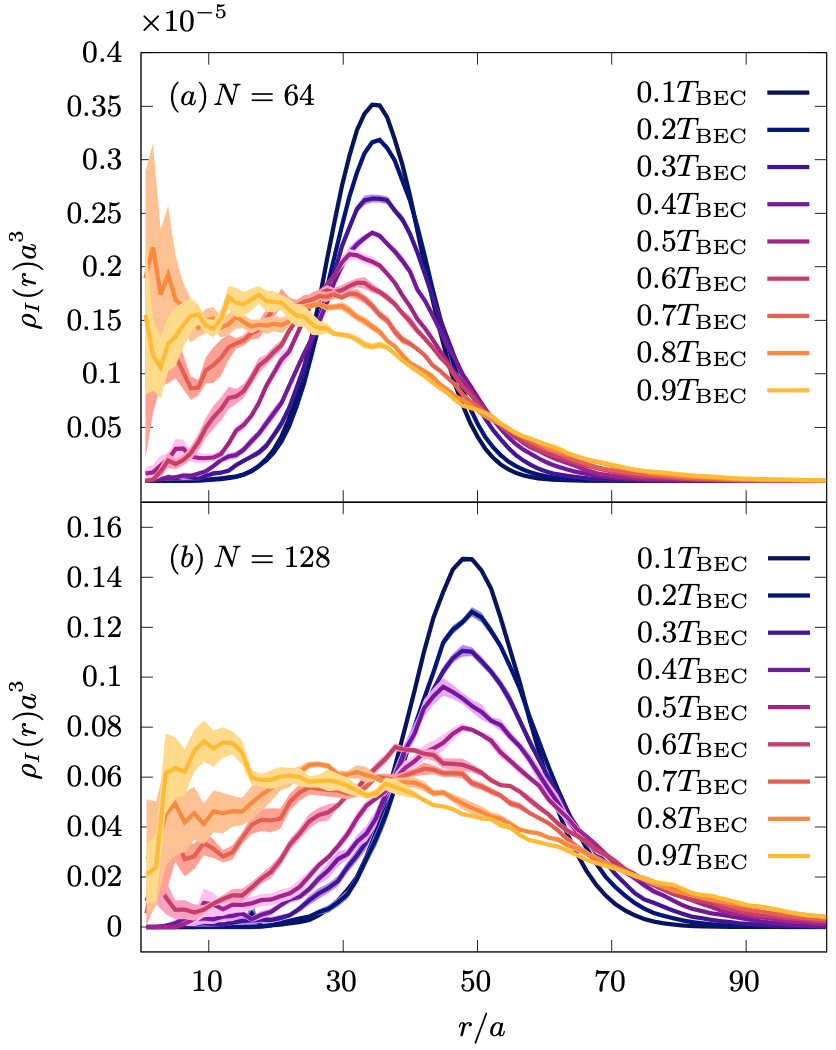}
	\caption{Density profiles of the impurity at different $N$ with $g_{BI}/g = 3.5$.}
    \label{fig:diff_N}
\end{figure}

In Fig.~\ref{fig:diff_N}, we show the differences in the 
density profiles between a system with $N=64$ and another with $N=128$ 
particles (keeping constant the harmonic confinement). Notice that the impurity 
moves to further distances when we increase $N$, as more particles in the bath 
are expelling it from the center (see Eq.\eqref{eq:radius_imp} where we show 
the dependence of the position of the impurity as a function of $N$). However, 
we do not notice a significance change in the mixing temperature $T_m$. 

\bibliography{manuscript}

\begin{thebibliography}{49}%
\makeatletter
\providecommand \@ifxundefined [1]{%
 \@ifx{#1\undefined}
}%
\providecommand \@ifnum [1]{%
 \ifnum #1\expandafter \@firstoftwo
 \else \expandafter \@secondoftwo
 \fi
}%
\providecommand \@ifx [1]{%
 \ifx #1\expandafter \@firstoftwo
 \else \expandafter \@secondoftwo
 \fi
}%
\providecommand \natexlab [1]{#1}%
\providecommand \enquote  [1]{``#1''}%
\providecommand \bibnamefont  [1]{#1}%
\providecommand \bibfnamefont [1]{#1}%
\providecommand \citenamefont [1]{#1}%
\providecommand \href@noop [0]{\@secondoftwo}%
\providecommand \href [0]{\begingroup \@sanitize@url \@href}%
\providecommand \@href[1]{\@@startlink{#1}\@@href}%
\providecommand \@@href[1]{\endgroup#1\@@endlink}%
\providecommand \@sanitize@url [0]{\catcode `\\12\catcode `\$12\catcode `\&12\catcode `\#12\catcode `\^12\catcode `\_12\catcode `\%12\relax}%
\providecommand \@@startlink[1]{}%
\providecommand \@@endlink[0]{}%
\providecommand \url  [0]{\begingroup\@sanitize@url \@url }%
\providecommand \@url [1]{\endgroup\@href {#1}{\urlprefix }}%
\providecommand \urlprefix  [0]{URL }%
\providecommand \Eprint [0]{\href }%
\providecommand \doibase [0]{https://doi.org/}%
\providecommand \selectlanguage [0]{\@gobble}%
\providecommand \bibinfo  [0]{\@secondoftwo}%
\providecommand \bibfield  [0]{\@secondoftwo}%
\providecommand \translation [1]{[#1]}%
\providecommand \BibitemOpen [0]{}%
\providecommand \bibitemStop [0]{}%
\providecommand \bibitemNoStop [0]{.\EOS\space}%
\providecommand \EOS [0]{\spacefactor3000\relax}%
\providecommand \BibitemShut  [1]{\csname bibitem#1\endcsname}%
\let\auto@bib@innerbib\@empty
\bibitem [{\citenamefont {Rath}\ and\ \citenamefont {Schmidt}(2013)}]{Rath2013}%
  \BibitemOpen
  \bibfield  {author} {\bibinfo {author} {\bibfnamefont {S.~P.}\ \bibnamefont {Rath}}\ and\ \bibinfo {author} {\bibfnamefont {R.}~\bibnamefont {Schmidt}},\ }\bibfield  {title} {\bibinfo {title} {{F}ield-theoretical study of the {B}ose polaron},\ }\href {https://doi.org/10.1103/PhysRevA.88.053632} {\bibfield  {journal} {\bibinfo  {journal} {Phys. Rev. A}\ }\textbf {\bibinfo {volume} {88}},\ \bibinfo {pages} {053632} (\bibinfo {year} {2013})}\BibitemShut {NoStop}%
\bibitem [{\citenamefont {Li}\ and\ \citenamefont {Das~Sarma}(2014)}]{Li2014}%
  \BibitemOpen
  \bibfield  {author} {\bibinfo {author} {\bibfnamefont {W.}~\bibnamefont {Li}}\ and\ \bibinfo {author} {\bibfnamefont {S.}~\bibnamefont {Das~Sarma}},\ }\bibfield  {title} {\bibinfo {title} {{V}ariational study of polarons in {B}ose-{E}instein condensates},\ }\href {https://doi.org/10.1103/PhysRevA.90.013618} {\bibfield  {journal} {\bibinfo  {journal} {Phys. Rev. A}\ }\textbf {\bibinfo {volume} {90}},\ \bibinfo {pages} {013618} (\bibinfo {year} {2014})}\BibitemShut {NoStop}%
\bibitem [{\citenamefont {Grusdt}\ \emph {et~al.}(2015)\citenamefont {Grusdt}, \citenamefont {Shchadilova}, \citenamefont {Rubtsov},\ and\ \citenamefont {Demler}}]{Grusdt2015}%
  \BibitemOpen
  \bibfield  {author} {\bibinfo {author} {\bibfnamefont {F.}~\bibnamefont {Grusdt}}, \bibinfo {author} {\bibfnamefont {Y.~E.}\ \bibnamefont {Shchadilova}}, \bibinfo {author} {\bibfnamefont {A.~N.}\ \bibnamefont {Rubtsov}},\ and\ \bibinfo {author} {\bibfnamefont {E.}~\bibnamefont {Demler}},\ }\bibfield  {title} {\bibinfo {title} {{R}enormalization group approach to the {F}r{\"o}hlich polaron model: application to impurity-{B}{E}{C} problem},\ }\href {https://doi.org/10.1038/srep12124} {\bibfield  {journal} {\bibinfo  {journal} {Scientific Reports}\ }\textbf {\bibinfo {volume} {5}},\ \bibinfo {pages} {12124} (\bibinfo {year} {2015})}\BibitemShut {NoStop}%
\bibitem [{\citenamefont {Ardila}\ and\ \citenamefont {Giorgini}(2015)}]{Ardila2015}%
  \BibitemOpen
  \bibfield  {author} {\bibinfo {author} {\bibfnamefont {L.~A.~P.}\ \bibnamefont {Ardila}}\ and\ \bibinfo {author} {\bibfnamefont {S.}~\bibnamefont {Giorgini}},\ }\bibfield  {title} {\bibinfo {title} {{I}mpurity in a {B}ose-{E}instein condensate: {S}tudy of the attractive and repulsive branch using quantum {M}onte {C}arlo methods},\ }\href {https://doi.org/10.1103/PhysRevA.92.033612} {\bibfield  {journal} {\bibinfo  {journal} {Phys. Rev. A}\ }\textbf {\bibinfo {volume} {92}},\ \bibinfo {pages} {033612} (\bibinfo {year} {2015})}\BibitemShut {NoStop}%
\bibitem [{\citenamefont {Volosniev}\ \emph {et~al.}(2015)\citenamefont {Volosniev}, \citenamefont {Hammer},\ and\ \citenamefont {Zinner}}]{Volosniev2015}%
  \BibitemOpen
  \bibfield  {author} {\bibinfo {author} {\bibfnamefont {A.~G.}\ \bibnamefont {Volosniev}}, \bibinfo {author} {\bibfnamefont {H.-W.}\ \bibnamefont {Hammer}},\ and\ \bibinfo {author} {\bibfnamefont {N.~T.}\ \bibnamefont {Zinner}},\ }\bibfield  {title} {\bibinfo {title} {{R}eal-time dynamics of an impurity in an ideal {B}ose gas in a trap},\ }\href {https://doi.org/10.1103/PhysRevA.92.023623} {\bibfield  {journal} {\bibinfo  {journal} {Phys. Rev. A}\ }\textbf {\bibinfo {volume} {92}},\ \bibinfo {pages} {023623} (\bibinfo {year} {2015})}\BibitemShut {NoStop}%
\bibitem [{\citenamefont {Levinsen}\ \emph {et~al.}(2015)\citenamefont {Levinsen}, \citenamefont {Parish},\ and\ \citenamefont {Bruun}}]{Levinsen2015}%
  \BibitemOpen
  \bibfield  {author} {\bibinfo {author} {\bibfnamefont {J.}~\bibnamefont {Levinsen}}, \bibinfo {author} {\bibfnamefont {M.~M.}\ \bibnamefont {Parish}},\ and\ \bibinfo {author} {\bibfnamefont {G.~M.}\ \bibnamefont {Bruun}},\ }\bibfield  {title} {\bibinfo {title} {{I}mpurity in a {B}ose-{E}instein {C}ondensate and the {E}fimov {E}ffect},\ }\href {https://doi.org/10.1103/PhysRevLett.115.125302} {\bibfield  {journal} {\bibinfo  {journal} {Phys. Rev. Lett.}\ }\textbf {\bibinfo {volume} {115}},\ \bibinfo {pages} {125302} (\bibinfo {year} {2015})}\BibitemShut {NoStop}%
\bibitem [{\citenamefont {Shchadilova}\ \emph {et~al.}(2016)\citenamefont {Shchadilova}, \citenamefont {Schmidt}, \citenamefont {Grusdt},\ and\ \citenamefont {Demler}}]{Shchadilova2016}%
  \BibitemOpen
  \bibfield  {author} {\bibinfo {author} {\bibfnamefont {Y.~E.}\ \bibnamefont {Shchadilova}}, \bibinfo {author} {\bibfnamefont {R.}~\bibnamefont {Schmidt}}, \bibinfo {author} {\bibfnamefont {F.}~\bibnamefont {Grusdt}},\ and\ \bibinfo {author} {\bibfnamefont {E.}~\bibnamefont {Demler}},\ }\bibfield  {title} {\bibinfo {title} {{Q}uantum {D}ynamics of {U}ltracold {B}ose {P}olarons},\ }\href {https://doi.org/10.1103/PhysRevLett.117.113002} {\bibfield  {journal} {\bibinfo  {journal} {Phys. Rev. Lett.}\ }\textbf {\bibinfo {volume} {117}},\ \bibinfo {pages} {113002} (\bibinfo {year} {2016})}\BibitemShut {NoStop}%
\bibitem [{\citenamefont {J\o{}rgensen}\ \emph {et~al.}(2016)\citenamefont {J\o{}rgensen}, \citenamefont {Wacker}, \citenamefont {Skalmstang}, \citenamefont {Parish}, \citenamefont {Levinsen}, \citenamefont {Christensen}, \citenamefont {Bruun},\ and\ \citenamefont {Arlt}}]{Jorgensen2016}%
  \BibitemOpen
  \bibfield  {author} {\bibinfo {author} {\bibfnamefont {N.~B.}\ \bibnamefont {J\o{}rgensen}}, \bibinfo {author} {\bibfnamefont {L.}~\bibnamefont {Wacker}}, \bibinfo {author} {\bibfnamefont {K.~T.}\ \bibnamefont {Skalmstang}}, \bibinfo {author} {\bibfnamefont {M.~M.}\ \bibnamefont {Parish}}, \bibinfo {author} {\bibfnamefont {J.}~\bibnamefont {Levinsen}}, \bibinfo {author} {\bibfnamefont {R.~S.}\ \bibnamefont {Christensen}}, \bibinfo {author} {\bibfnamefont {G.~M.}\ \bibnamefont {Bruun}},\ and\ \bibinfo {author} {\bibfnamefont {J.~J.}\ \bibnamefont {Arlt}},\ }\bibfield  {title} {\bibinfo {title} {{O}bservation of {A}ttractive and {R}epulsive {P}olarons in a {B}ose-{E}instein {C}ondensate},\ }\href {https://doi.org/10.1103/PhysRevLett.117.055302} {\bibfield  {journal} {\bibinfo  {journal} {Phys. Rev. Lett.}\ }\textbf {\bibinfo {volume} {117}},\ \bibinfo {pages} {055302} (\bibinfo {year} {2016})}\BibitemShut {NoStop}%
\bibitem [{\citenamefont {Hu}\ \emph {et~al.}(2016)\citenamefont {Hu}, \citenamefont {Van~de Graaff}, \citenamefont {Kedar}, \citenamefont {Corson}, \citenamefont {Cornell},\ and\ \citenamefont {Jin}}]{Hu2016}%
  \BibitemOpen
  \bibfield  {author} {\bibinfo {author} {\bibfnamefont {M.-G.}\ \bibnamefont {Hu}}, \bibinfo {author} {\bibfnamefont {M.~J.}\ \bibnamefont {Van~de Graaff}}, \bibinfo {author} {\bibfnamefont {D.}~\bibnamefont {Kedar}}, \bibinfo {author} {\bibfnamefont {J.~P.}\ \bibnamefont {Corson}}, \bibinfo {author} {\bibfnamefont {E.~A.}\ \bibnamefont {Cornell}},\ and\ \bibinfo {author} {\bibfnamefont {D.~S.}\ \bibnamefont {Jin}},\ }\bibfield  {title} {\bibinfo {title} {{B}ose {P}olarons in the {S}trongly {I}nteracting {R}egime},\ }\href {https://doi.org/10.1103/PhysRevLett.117.055301} {\bibfield  {journal} {\bibinfo  {journal} {Phys. Rev. Lett.}\ }\textbf {\bibinfo {volume} {117}},\ \bibinfo {pages} {055301} (\bibinfo {year} {2016})}\BibitemShut {NoStop}%
\bibitem [{\citenamefont {Ardila}\ and\ \citenamefont {Giorgini}(2016)}]{Ardila2016}%
  \BibitemOpen
  \bibfield  {author} {\bibinfo {author} {\bibfnamefont {L.~A.~P.}\ \bibnamefont {Ardila}}\ and\ \bibinfo {author} {\bibfnamefont {S.}~\bibnamefont {Giorgini}},\ }\bibfield  {title} {\bibinfo {title} {{B}ose polaron problem: {E}ffect of mass imbalance on binding energy},\ }\href {https://doi.org/10.1103/PhysRevA.94.063640} {\bibfield  {journal} {\bibinfo  {journal} {Phys. Rev. A}\ }\textbf {\bibinfo {volume} {94}},\ \bibinfo {pages} {063640} (\bibinfo {year} {2016})}\BibitemShut {NoStop}%
\bibitem [{\citenamefont {Sun}\ \emph {et~al.}(2017)\citenamefont {Sun}, \citenamefont {Zhai},\ and\ \citenamefont {Cui}}]{Sun2017}%
  \BibitemOpen
  \bibfield  {author} {\bibinfo {author} {\bibfnamefont {M.}~\bibnamefont {Sun}}, \bibinfo {author} {\bibfnamefont {H.}~\bibnamefont {Zhai}},\ and\ \bibinfo {author} {\bibfnamefont {X.}~\bibnamefont {Cui}},\ }\bibfield  {title} {\bibinfo {title} {{V}isualizing the {E}fimov {C}orrelation in {B}ose {P}olarons},\ }\href {https://doi.org/10.1103/PhysRevLett.119.013401} {\bibfield  {journal} {\bibinfo  {journal} {Phys. Rev. Lett.}\ }\textbf {\bibinfo {volume} {119}},\ \bibinfo {pages} {013401} (\bibinfo {year} {2017})}\BibitemShut {NoStop}%
\bibitem [{\citenamefont {Yoshida}\ \emph {et~al.}(2018)\citenamefont {Yoshida}, \citenamefont {Endo}, \citenamefont {Levinsen},\ and\ \citenamefont {Parish}}]{Yoshida2018}%
  \BibitemOpen
  \bibfield  {author} {\bibinfo {author} {\bibfnamefont {S.~M.}\ \bibnamefont {Yoshida}}, \bibinfo {author} {\bibfnamefont {S.}~\bibnamefont {Endo}}, \bibinfo {author} {\bibfnamefont {J.}~\bibnamefont {Levinsen}},\ and\ \bibinfo {author} {\bibfnamefont {M.~M.}\ \bibnamefont {Parish}},\ }\bibfield  {title} {\bibinfo {title} {{U}niversality of an {I}mpurity in a {B}ose-{E}instein {C}ondensate},\ }\href {https://doi.org/10.1103/PhysRevX.8.011024} {\bibfield  {journal} {\bibinfo  {journal} {Phys. Rev. X}\ }\textbf {\bibinfo {volume} {8}},\ \bibinfo {pages} {011024} (\bibinfo {year} {2018})}\BibitemShut {NoStop}%
\bibitem [{\citenamefont {Van~Loon}\ \emph {et~al.}(2018)\citenamefont {Van~Loon}, \citenamefont {Casteels},\ and\ \citenamefont {Tempere}}]{Loon2018}%
  \BibitemOpen
  \bibfield  {author} {\bibinfo {author} {\bibfnamefont {S.}~\bibnamefont {Van~Loon}}, \bibinfo {author} {\bibfnamefont {W.}~\bibnamefont {Casteels}},\ and\ \bibinfo {author} {\bibfnamefont {J.}~\bibnamefont {Tempere}},\ }\bibfield  {title} {\bibinfo {title} {{G}round-state properties of interacting {B}ose polarons},\ }\href {https://doi.org/10.1103/PhysRevA.98.063631} {\bibfield  {journal} {\bibinfo  {journal} {Phys. Rev. A}\ }\textbf {\bibinfo {volume} {98}},\ \bibinfo {pages} {063631} (\bibinfo {year} {2018})}\BibitemShut {NoStop}%
\bibitem [{\citenamefont {Mistakidis}\ \emph {et~al.}(2019)\citenamefont {Mistakidis}, \citenamefont {Katsimiga}, \citenamefont {Koutentakis}, \citenamefont {Busch},\ and\ \citenamefont {Schmelcher}}]{Mistakidis2019}%
  \BibitemOpen
  \bibfield  {author} {\bibinfo {author} {\bibfnamefont {S.~I.}\ \bibnamefont {Mistakidis}}, \bibinfo {author} {\bibfnamefont {G.~C.}\ \bibnamefont {Katsimiga}}, \bibinfo {author} {\bibfnamefont {G.~M.}\ \bibnamefont {Koutentakis}}, \bibinfo {author} {\bibfnamefont {T.}~\bibnamefont {Busch}},\ and\ \bibinfo {author} {\bibfnamefont {P.}~\bibnamefont {Schmelcher}},\ }\bibfield  {title} {\bibinfo {title} {{Q}uench {D}ynamics and {O}rthogonality {C}atastrophe of {B}ose {P}olarons},\ }\href {https://doi.org/10.1103/PhysRevLett.122.183001} {\bibfield  {journal} {\bibinfo  {journal} {Phys. Rev. Lett.}\ }\textbf {\bibinfo {volume} {122}},\ \bibinfo {pages} {183001} (\bibinfo {year} {2019})}\BibitemShut {NoStop}%
\bibitem [{\citenamefont {Ichmoukhamedov}\ and\ \citenamefont {Tempere}(2019)}]{Ichmoukhamedov2019}%
  \BibitemOpen
  \bibfield  {author} {\bibinfo {author} {\bibfnamefont {T.}~\bibnamefont {Ichmoukhamedov}}\ and\ \bibinfo {author} {\bibfnamefont {J.}~\bibnamefont {Tempere}},\ }\bibfield  {title} {\bibinfo {title} {{F}eynman path-integral treatment of the {B}ose polaron beyond the {F}r\"ohlich model},\ }\href {https://doi.org/10.1103/PhysRevA.100.043605} {\bibfield  {journal} {\bibinfo  {journal} {Phys. Rev. A}\ }\textbf {\bibinfo {volume} {100}},\ \bibinfo {pages} {043605} (\bibinfo {year} {2019})}\BibitemShut {NoStop}%
\bibitem [{\citenamefont {Drescher}\ \emph {et~al.}(2020)\citenamefont {Drescher}, \citenamefont {Salmhofer},\ and\ \citenamefont {Enss}}]{Drescher2020}%
  \BibitemOpen
  \bibfield  {author} {\bibinfo {author} {\bibfnamefont {M.}~\bibnamefont {Drescher}}, \bibinfo {author} {\bibfnamefont {M.}~\bibnamefont {Salmhofer}},\ and\ \bibinfo {author} {\bibfnamefont {T.}~\bibnamefont {Enss}},\ }\bibfield  {title} {\bibinfo {title} {{T}heory of a resonantly interacting impurity in a {B}ose-{E}instein condensate},\ }\href {https://doi.org/10.1103/PhysRevResearch.2.032011} {\bibfield  {journal} {\bibinfo  {journal} {Phys. Rev. Res.}\ }\textbf {\bibinfo {volume} {2}},\ \bibinfo {pages} {032011} (\bibinfo {year} {2020})}\BibitemShut {NoStop}%
\bibitem [{\citenamefont {Levinsen}\ \emph {et~al.}(2021)\citenamefont {Levinsen}, \citenamefont {Ardila}, \citenamefont {Yoshida},\ and\ \citenamefont {Parish}}]{Levinsen2021}%
  \BibitemOpen
  \bibfield  {author} {\bibinfo {author} {\bibfnamefont {J.}~\bibnamefont {Levinsen}}, \bibinfo {author} {\bibfnamefont {L.~A.~P.}\ \bibnamefont {Ardila}}, \bibinfo {author} {\bibfnamefont {S.~M.}\ \bibnamefont {Yoshida}},\ and\ \bibinfo {author} {\bibfnamefont {M.~M.}\ \bibnamefont {Parish}},\ }\bibfield  {title} {\bibinfo {title} {{Q}uantum behavior of a heavy impurity strongly coupled to a {B}ose gas},\ }\href {https://doi.org/10.1103/PhysRevLett.127.033401} {\bibfield  {journal} {\bibinfo  {journal} {Phys. Rev. Lett.}\ }\textbf {\bibinfo {volume} {127}},\ \bibinfo {pages} {033401} (\bibinfo {year} {2021})}\BibitemShut {NoStop}%
\bibitem [{\citenamefont {Massignan}\ \emph {et~al.}(2021)\citenamefont {Massignan}, \citenamefont {Yegovtsev},\ and\ \citenamefont {Gurarie}}]{Massignan2021}%
  \BibitemOpen
  \bibfield  {author} {\bibinfo {author} {\bibfnamefont {P.}~\bibnamefont {Massignan}}, \bibinfo {author} {\bibfnamefont {N.}~\bibnamefont {Yegovtsev}},\ and\ \bibinfo {author} {\bibfnamefont {V.}~\bibnamefont {Gurarie}},\ }\bibfield  {title} {\bibinfo {title} {{U}niversal aspects of a strongly interacting impurity in a dilute {B}ose condensate},\ }\href {https://doi.org/10.1103/PhysRevLett.126.123403} {\bibfield  {journal} {\bibinfo  {journal} {Phys. Rev. Lett.}\ }\textbf {\bibinfo {volume} {126}},\ \bibinfo {pages} {123403} (\bibinfo {year} {2021})}\BibitemShut {NoStop}%
\bibitem [{\citenamefont {Isaule}\ \emph {et~al.}(2021)\citenamefont {Isaule}, \citenamefont {Morera}, \citenamefont {Massignan},\ and\ \citenamefont {Juli\'a-D\'{\i}az}}]{Isaule2021}%
  \BibitemOpen
  \bibfield  {author} {\bibinfo {author} {\bibfnamefont {F.}~\bibnamefont {Isaule}}, \bibinfo {author} {\bibfnamefont {I.}~\bibnamefont {Morera}}, \bibinfo {author} {\bibfnamefont {P.}~\bibnamefont {Massignan}},\ and\ \bibinfo {author} {\bibfnamefont {B.}~\bibnamefont {Juli\'a-D\'{\i}az}},\ }\bibfield  {title} {\bibinfo {title} {{R}enormalization-group study of {B}ose polarons},\ }\href {https://doi.org/10.1103/PhysRevA.104.023317} {\bibfield  {journal} {\bibinfo  {journal} {Phys. Rev. A}\ }\textbf {\bibinfo {volume} {104}},\ \bibinfo {pages} {023317} (\bibinfo {year} {2021})}\BibitemShut {NoStop}%
\bibitem [{\citenamefont {Christianen}\ \emph {et~al.}(2022{\natexlab{a}})\citenamefont {Christianen}, \citenamefont {Cirac},\ and\ \citenamefont {Schmidt}}]{Christianen2022}%
  \BibitemOpen
  \bibfield  {author} {\bibinfo {author} {\bibfnamefont {A.}~\bibnamefont {Christianen}}, \bibinfo {author} {\bibfnamefont {J.~I.}\ \bibnamefont {Cirac}},\ and\ \bibinfo {author} {\bibfnamefont {R.}~\bibnamefont {Schmidt}},\ }\bibfield  {title} {\bibinfo {title} {{C}hemistry of a light impurity in a {B}ose-{E}instein condensate},\ }\href {https://doi.org/10.1103/PhysRevLett.128.183401} {\bibfield  {journal} {\bibinfo  {journal} {Phys. Rev. Lett.}\ }\textbf {\bibinfo {volume} {128}},\ \bibinfo {pages} {183401} (\bibinfo {year} {2022}{\natexlab{a}})}\BibitemShut {NoStop}%
\bibitem [{\citenamefont {Christianen}\ \emph {et~al.}(2022{\natexlab{b}})\citenamefont {Christianen}, \citenamefont {Cirac},\ and\ \citenamefont {Schmidt}}]{Christianen2022PRA}%
  \BibitemOpen
  \bibfield  {author} {\bibinfo {author} {\bibfnamefont {A.}~\bibnamefont {Christianen}}, \bibinfo {author} {\bibfnamefont {J.~I.}\ \bibnamefont {Cirac}},\ and\ \bibinfo {author} {\bibfnamefont {R.}~\bibnamefont {Schmidt}},\ }\bibfield  {title} {\bibinfo {title} {{B}ose polaron and the {E}fimov effect: {A} {G}aussian-state approach},\ }\href {https://doi.org/10.1103/PhysRevA.105.053302} {\bibfield  {journal} {\bibinfo  {journal} {Phys. Rev. A}\ }\textbf {\bibinfo {volume} {105}},\ \bibinfo {pages} {053302} (\bibinfo {year} {2022}{\natexlab{b}})}\BibitemShut {NoStop}%
\bibitem [{\citenamefont {Skou}\ \emph {et~al.}(2022)\citenamefont {Skou}, \citenamefont {Nielsen}, \citenamefont {Skov}, \citenamefont {Morgen}, \citenamefont {J\o{}rgensen}, \citenamefont {Camacho-Guardian}, \citenamefont {Pohl}, \citenamefont {Bruun},\ and\ \citenamefont {Arlt}}]{Skou2022}%
  \BibitemOpen
  \bibfield  {author} {\bibinfo {author} {\bibfnamefont {M.~G.}\ \bibnamefont {Skou}}, \bibinfo {author} {\bibfnamefont {K.~K.}\ \bibnamefont {Nielsen}}, \bibinfo {author} {\bibfnamefont {T.~G.}\ \bibnamefont {Skov}}, \bibinfo {author} {\bibfnamefont {A.~M.}\ \bibnamefont {Morgen}}, \bibinfo {author} {\bibfnamefont {N.~B.}\ \bibnamefont {J\o{}rgensen}}, \bibinfo {author} {\bibfnamefont {A.}~\bibnamefont {Camacho-Guardian}}, \bibinfo {author} {\bibfnamefont {T.}~\bibnamefont {Pohl}}, \bibinfo {author} {\bibfnamefont {G.~M.}\ \bibnamefont {Bruun}},\ and\ \bibinfo {author} {\bibfnamefont {J.~J.}\ \bibnamefont {Arlt}},\ }\bibfield  {title} {\bibinfo {title} {{L}ife and death of the {B}ose polaron},\ }\href {https://doi.org/10.1103/PhysRevResearch.4.043093} {\bibfield  {journal} {\bibinfo  {journal} {Phys. Rev. Res.}\ }\textbf {\bibinfo {volume} {4}},\ \bibinfo {pages} {043093} (\bibinfo {year} {2022})}\BibitemShut {NoStop}%
\bibitem [{\citenamefont {Tempere}\ \emph {et~al.}(2009)\citenamefont {Tempere}, \citenamefont {Casteels}, \citenamefont {Oberthaler}, \citenamefont {Knoop}, \citenamefont {Timmermans},\ and\ \citenamefont {Devreese}}]{Tempere2009}%
  \BibitemOpen
  \bibfield  {author} {\bibinfo {author} {\bibfnamefont {J.}~\bibnamefont {Tempere}}, \bibinfo {author} {\bibfnamefont {W.}~\bibnamefont {Casteels}}, \bibinfo {author} {\bibfnamefont {M.~K.}\ \bibnamefont {Oberthaler}}, \bibinfo {author} {\bibfnamefont {S.}~\bibnamefont {Knoop}}, \bibinfo {author} {\bibfnamefont {E.}~\bibnamefont {Timmermans}},\ and\ \bibinfo {author} {\bibfnamefont {J.~T.}\ \bibnamefont {Devreese}},\ }\bibfield  {title} {\bibinfo {title} {{F}eynman path-integral treatment of the {B}{E}{C}-impurity polaron},\ }\href {https://doi.org/10.1103/PhysRevB.80.184504} {\bibfield  {journal} {\bibinfo  {journal} {Phys. Rev. B}\ }\textbf {\bibinfo {volume} {80}},\ \bibinfo {pages} {184504} (\bibinfo {year} {2009})}\BibitemShut {NoStop}%
\bibitem [{\citenamefont {Guenther}\ \emph {et~al.}(2018)\citenamefont {Guenther}, \citenamefont {Massignan}, \citenamefont {Lewenstein},\ and\ \citenamefont {Bruun}}]{Guenther2018}%
  \BibitemOpen
  \bibfield  {author} {\bibinfo {author} {\bibfnamefont {N.-E.}\ \bibnamefont {Guenther}}, \bibinfo {author} {\bibfnamefont {P.}~\bibnamefont {Massignan}}, \bibinfo {author} {\bibfnamefont {M.}~\bibnamefont {Lewenstein}},\ and\ \bibinfo {author} {\bibfnamefont {G.~M.}\ \bibnamefont {Bruun}},\ }\bibfield  {title} {\bibinfo {title} {{B}ose {P}olarons at {F}inite {T}emperature and {S}trong {C}oupling},\ }\href {https://doi.org/10.1103/PhysRevLett.120.050405} {\bibfield  {journal} {\bibinfo  {journal} {Phys. Rev. Lett.}\ }\textbf {\bibinfo {volume} {120}},\ \bibinfo {pages} {050405} (\bibinfo {year} {2018})}\BibitemShut {NoStop}%
\bibitem [{\citenamefont {Pastukhov}(2018)}]{Pastukhov2018}%
  \BibitemOpen
  \bibfield  {author} {\bibinfo {author} {\bibfnamefont {V.}~\bibnamefont {Pastukhov}},\ }\bibfield  {title} {\bibinfo {title} {{P}olaron in the dilute critical {B}ose condensate},\ }\href {https://doi.org/10.1088/1751-8121/aab9c1} {\bibfield  {journal} {\bibinfo  {journal} {Journal of Physics A: Mathematical and Theoretical}\ }\textbf {\bibinfo {volume} {51}},\ \bibinfo {pages} {195003} (\bibinfo {year} {2018})}\BibitemShut {NoStop}%
\bibitem [{\citenamefont {Field}\ \emph {et~al.}(2020)\citenamefont {Field}, \citenamefont {Levinsen},\ and\ \citenamefont {Parish}}]{Field2020}%
  \BibitemOpen
  \bibfield  {author} {\bibinfo {author} {\bibfnamefont {B.}~\bibnamefont {Field}}, \bibinfo {author} {\bibfnamefont {J.}~\bibnamefont {Levinsen}},\ and\ \bibinfo {author} {\bibfnamefont {M.~M.}\ \bibnamefont {Parish}},\ }\bibfield  {title} {\bibinfo {title} {{F}ate of the {B}ose polaron at finite temperature},\ }\href {https://doi.org/10.1103/PhysRevA.101.013623} {\bibfield  {journal} {\bibinfo  {journal} {Phys. Rev. A}\ }\textbf {\bibinfo {volume} {101}},\ \bibinfo {pages} {013623} (\bibinfo {year} {2020})}\BibitemShut {NoStop}%
\bibitem [{\citenamefont {Yan}\ \emph {et~al.}(2020)\citenamefont {Yan}, \citenamefont {Ni}, \citenamefont {Robens},\ and\ \citenamefont {Zwierlein}}]{Zoe2020}%
  \BibitemOpen
  \bibfield  {author} {\bibinfo {author} {\bibfnamefont {Z.~Z.}\ \bibnamefont {Yan}}, \bibinfo {author} {\bibfnamefont {Y.}~\bibnamefont {Ni}}, \bibinfo {author} {\bibfnamefont {C.}~\bibnamefont {Robens}},\ and\ \bibinfo {author} {\bibfnamefont {M.~W.}\ \bibnamefont {Zwierlein}},\ }\bibfield  {title} {\bibinfo {title} {{B}ose polarons near quantum criticality},\ }\href {https://doi.org/10.1126/science.aax5850} {\bibfield  {journal} {\bibinfo  {journal} {Science}\ }\textbf {\bibinfo {volume} {368}},\ \bibinfo {pages} {190} (\bibinfo {year} {2020})},\ \Eprint {https://arxiv.org/abs/https://www.science.org/doi/pdf/10.1126/science.aax5850} {https://www.science.org/doi/pdf/10.1126/science.aax5850} \BibitemShut {NoStop}%
\bibitem [{\citenamefont {Pascual}\ and\ \citenamefont {Boronat}(2021)}]{Pascual2021}%
  \BibitemOpen
  \bibfield  {author} {\bibinfo {author} {\bibfnamefont {G.}~\bibnamefont {Pascual}}\ and\ \bibinfo {author} {\bibfnamefont {J.}~\bibnamefont {Boronat}},\ }\bibfield  {title} {\bibinfo {title} {{Q}uasiparticle nature of the {B}ose polaron at finite temperature},\ }\href {https://doi.org/10.1103/PhysRevLett.127.205301} {\bibfield  {journal} {\bibinfo  {journal} {Phys. Rev. Lett.}\ }\textbf {\bibinfo {volume} {127}},\ \bibinfo {pages} {205301} (\bibinfo {year} {2021})}\BibitemShut {NoStop}%
\bibitem [{\citenamefont {Mehboudi}\ \emph {et~al.}(2019)\citenamefont {Mehboudi}, \citenamefont {Lampo}, \citenamefont {Charalambous}, \citenamefont {Correa}, \citenamefont {Garc\'{\i}a-March},\ and\ \citenamefont {Lewenstein}}]{Mehboudi2019}%
  \BibitemOpen
  \bibfield  {author} {\bibinfo {author} {\bibfnamefont {M.}~\bibnamefont {Mehboudi}}, \bibinfo {author} {\bibfnamefont {A.}~\bibnamefont {Lampo}}, \bibinfo {author} {\bibfnamefont {C.}~\bibnamefont {Charalambous}}, \bibinfo {author} {\bibfnamefont {L.~A.}\ \bibnamefont {Correa}}, \bibinfo {author} {\bibfnamefont {M.~A.}\ \bibnamefont {Garc\'{\i}a-March}},\ and\ \bibinfo {author} {\bibfnamefont {M.}~\bibnamefont {Lewenstein}},\ }\bibfield  {title} {\bibinfo {title} {{U}sing polarons for sub-nk quantum nondemolition thermometry in a {B}ose-{E}instein condensate},\ }\href {https://doi.org/10.1103/PhysRevLett.122.030403} {\bibfield  {journal} {\bibinfo  {journal} {Phys. Rev. Lett.}\ }\textbf {\bibinfo {volume} {122}},\ \bibinfo {pages} {030403} (\bibinfo {year} {2019})}\BibitemShut {NoStop}%
\bibitem [{\citenamefont {Ao}\ and\ \citenamefont {Chui}(1998)}]{Ao1998}%
  \BibitemOpen
  \bibfield  {author} {\bibinfo {author} {\bibfnamefont {P.}~\bibnamefont {Ao}}\ and\ \bibinfo {author} {\bibfnamefont {S.~T.}\ \bibnamefont {Chui}},\ }\bibfield  {title} {\bibinfo {title} {{B}inary {B}ose-{E}instein condensate mixtures in weakly and strongly segregated phases},\ }\href {https://doi.org/10.1103/PhysRevA.58.4836} {\bibfield  {journal} {\bibinfo  {journal} {Phys. Rev. A}\ }\textbf {\bibinfo {volume} {58}},\ \bibinfo {pages} {4836} (\bibinfo {year} {1998})}\BibitemShut {NoStop}%
\bibitem [{\citenamefont {Ota}\ \emph {et~al.}(2019)\citenamefont {Ota}, \citenamefont {Giorgini},\ and\ \citenamefont {Stringari}}]{Ota2019}%
  \BibitemOpen
  \bibfield  {author} {\bibinfo {author} {\bibfnamefont {M.}~\bibnamefont {Ota}}, \bibinfo {author} {\bibfnamefont {S.}~\bibnamefont {Giorgini}},\ and\ \bibinfo {author} {\bibfnamefont {S.}~\bibnamefont {Stringari}},\ }\bibfield  {title} {\bibinfo {title} {{M}agnetic phase transition in a mixture of two interacting superfluid {B}ose gases at finite temperature},\ }\href {https://doi.org/10.1103/PhysRevLett.123.075301} {\bibfield  {journal} {\bibinfo  {journal} {Phys. Rev. Lett.}\ }\textbf {\bibinfo {volume} {123}},\ \bibinfo {pages} {075301} (\bibinfo {year} {2019})}\BibitemShut {NoStop}%
\bibitem [{\citenamefont {Ota}\ and\ \citenamefont {Giorgini}(2020)}]{Ota2020}%
  \BibitemOpen
  \bibfield  {author} {\bibinfo {author} {\bibfnamefont {M.}~\bibnamefont {Ota}}\ and\ \bibinfo {author} {\bibfnamefont {S.}~\bibnamefont {Giorgini}},\ }\bibfield  {title} {\bibinfo {title} {{T}hermodynamics of dilute {B}ose gases: Beyond mean-field theory for binary mixtures of {B}ose-{E}instein condensates},\ }\href {https://doi.org/10.1103/PhysRevA.102.063303} {\bibfield  {journal} {\bibinfo  {journal} {Phys. Rev. A}\ }\textbf {\bibinfo {volume} {102}},\ \bibinfo {pages} {063303} (\bibinfo {year} {2020})}\BibitemShut {NoStop}%
\bibitem [{\citenamefont {Spada}\ \emph {et~al.}(2023)\citenamefont {Spada}, \citenamefont {Parisi}, \citenamefont {Pascual}, \citenamefont {Parker}, \citenamefont {Billam}, \citenamefont {Pilati}, \citenamefont {Boronat},\ and\ \citenamefont {Giorgini}}]{Spada2023}%
  \BibitemOpen
  \bibfield  {author} {\bibinfo {author} {\bibfnamefont {G.}~\bibnamefont {Spada}}, \bibinfo {author} {\bibfnamefont {L.}~\bibnamefont {Parisi}}, \bibinfo {author} {\bibfnamefont {G.}~\bibnamefont {Pascual}}, \bibinfo {author} {\bibfnamefont {N.~G.}\ \bibnamefont {Parker}}, \bibinfo {author} {\bibfnamefont {T.~P.}\ \bibnamefont {Billam}}, \bibinfo {author} {\bibfnamefont {S.}~\bibnamefont {Pilati}}, \bibinfo {author} {\bibfnamefont {J.}~\bibnamefont {Boronat}},\ and\ \bibinfo {author} {\bibfnamefont {S.}~\bibnamefont {Giorgini}},\ }\bibfield  {title} {\bibinfo {title} {{Phase separation in binary Bose mixtures at finite temperature}},\ }\href {https://doi.org/10.21468/SciPostPhys.15.4.171} {\bibfield  {journal} {\bibinfo  {journal} {SciPost Phys.}\ }\textbf {\bibinfo {volume} {15}},\ \bibinfo {pages} {171} (\bibinfo {year} {2023})}\BibitemShut {NoStop}%
\bibitem [{\citenamefont {Pascual}\ \emph {et~al.}(2023)\citenamefont {Pascual}, \citenamefont {Spada}, \citenamefont {Pilati}, \citenamefont {Giorgini},\ and\ \citenamefont {Boronat}}]{Pascual2023}%
  \BibitemOpen
  \bibfield  {author} {\bibinfo {author} {\bibfnamefont {G.}~\bibnamefont {Pascual}}, \bibinfo {author} {\bibfnamefont {G.}~\bibnamefont {Spada}}, \bibinfo {author} {\bibfnamefont {S.}~\bibnamefont {Pilati}}, \bibinfo {author} {\bibfnamefont {S.}~\bibnamefont {Giorgini}},\ and\ \bibinfo {author} {\bibfnamefont {J.}~\bibnamefont {Boronat}},\ }\bibfield  {title} {\bibinfo {title} {{T}hermally induced local imbalance in repulsive binary {B}ose mixtures},\ }\href {https://doi.org/10.1103/PhysRevResearch.5.L032041} {\bibfield  {journal} {\bibinfo  {journal} {Phys. Rev. Res.}\ }\textbf {\bibinfo {volume} {5}},\ \bibinfo {pages} {L032041} (\bibinfo {year} {2023})}\BibitemShut {NoStop}%
\bibitem [{\citenamefont {D\ifmmode~\check{z}\else \v{z}\fi{}elalija}\ \emph {et~al.}(2020)\citenamefont {D\ifmmode~\check{z}\else \v{z}\fi{}elalija}, \citenamefont {Cikojevi\ifmmode~\acute{c}\else \'{c}\fi{}}, \citenamefont {Boronat},\ and\ \citenamefont {Vranje\ifmmode \check{s}\else \v{s}\fi{} Marki\ifmmode~\acute{c}\else \'{c}\fi{}}}]{Dzelalija2020}%
  \BibitemOpen
  \bibfield  {author} {\bibinfo {author} {\bibfnamefont {K.}~\bibnamefont {D\ifmmode~\check{z}\else \v{z}\fi{}elalija}}, \bibinfo {author} {\bibfnamefont {V.}~\bibnamefont {Cikojevi\ifmmode~\acute{c}\else \'{c}\fi{}}}, \bibinfo {author} {\bibfnamefont {J.}~\bibnamefont {Boronat}},\ and\ \bibinfo {author} {\bibfnamefont {L.}~\bibnamefont {Vranje\ifmmode \check{s}\else \v{s}\fi{} Marki\ifmmode~\acute{c}\else \'{c}\fi{}}},\ }\bibfield  {title} {\bibinfo {title} {{T}rapped {B}ose-{B}ose mixtures at finite temperature: {A} quantum {M}onte {C}arlo approach},\ }\href {https://doi.org/10.1103/PhysRevA.102.063304} {\bibfield  {journal} {\bibinfo  {journal} {Phys. Rev. A}\ }\textbf {\bibinfo {volume} {102}},\ \bibinfo {pages} {063304} (\bibinfo {year} {2020})}\BibitemShut {NoStop}%
\bibitem [{\citenamefont {Pilati}\ \emph {et~al.}(2006)\citenamefont {Pilati}, \citenamefont {Sakkos}, \citenamefont {Boronat}, \citenamefont {Casulleras},\ and\ \citenamefont {Giorgini}}]{Pilati2006}%
  \BibitemOpen
  \bibfield  {author} {\bibinfo {author} {\bibfnamefont {S.}~\bibnamefont {Pilati}}, \bibinfo {author} {\bibfnamefont {K.}~\bibnamefont {Sakkos}}, \bibinfo {author} {\bibfnamefont {J.}~\bibnamefont {Boronat}}, \bibinfo {author} {\bibfnamefont {J.}~\bibnamefont {Casulleras}},\ and\ \bibinfo {author} {\bibfnamefont {S.}~\bibnamefont {Giorgini}},\ }\bibfield  {title} {\bibinfo {title} {{E}quation of state of an interacting {B}ose gas at finite temperature: {A} path-integral {M}onte {C}arlo study},\ }\href {https://doi.org/10.1103/PhysRevA.74.043621} {\bibfield  {journal} {\bibinfo  {journal} {Phys. Rev. A}\ }\textbf {\bibinfo {volume} {74}},\ \bibinfo {pages} {043621} (\bibinfo {year} {2006})}\BibitemShut {NoStop}%
\bibitem [{\citenamefont {Landau}\ and\ \citenamefont {Lifshitz}(1977)}]{Landau1977}%
  \BibitemOpen
  \bibfield  {author} {\bibinfo {author} {\bibfnamefont {L.~D.}\ \bibnamefont {Landau}}\ and\ \bibinfo {author} {\bibfnamefont {E.~M.}\ \bibnamefont {Lifshitz}},\ }\href@noop {} {\emph {\bibinfo {title} {{Q}uantum {M}echanics ({N}onrelativistic {T}heory)}}}\ (\bibinfo  {publisher} {{P}ergamon {P}ress},\ \bibinfo {address} {{O}xford},\ \bibinfo {year} {1977})\ p.\ \bibinfo {pages} {550}\BibitemShut {NoStop}%
\bibitem [{\citenamefont {Giorgini}\ \emph {et~al.}(1999)\citenamefont {Giorgini}, \citenamefont {Boronat},\ and\ \citenamefont {Casulleras}}]{Giorgini1999}%
  \BibitemOpen
  \bibfield  {author} {\bibinfo {author} {\bibfnamefont {S.}~\bibnamefont {Giorgini}}, \bibinfo {author} {\bibfnamefont {J.}~\bibnamefont {Boronat}},\ and\ \bibinfo {author} {\bibfnamefont {J.}~\bibnamefont {Casulleras}},\ }\bibfield  {title} {\bibinfo {title} {{G}round state of a homogeneous {B}ose gas: {A} diffusion {M}onte {C}arlo calculation},\ }\href {https://doi.org/10.1103/PhysRevA.60.5129} {\bibfield  {journal} {\bibinfo  {journal} {Phys. Rev. A}\ }\textbf {\bibinfo {volume} {60}},\ \bibinfo {pages} {5129} (\bibinfo {year} {1999})}\BibitemShut {NoStop}%
\bibitem [{\citenamefont {Ceperley}(1995)}]{Ceperley1995}%
  \BibitemOpen
  \bibfield  {author} {\bibinfo {author} {\bibfnamefont {D.~M.}\ \bibnamefont {Ceperley}},\ }\bibfield  {title} {\bibinfo {title} {{P}ath integrals in the theory of condensed {H}elium},\ }\href {https://doi.org/10.1103/RevModPhys.67.279} {\bibfield  {journal} {\bibinfo  {journal} {Rev. Mod. Phys.}\ }\textbf {\bibinfo {volume} {67}},\ \bibinfo {pages} {279} (\bibinfo {year} {1995})}\BibitemShut {NoStop}%
\bibitem [{\citenamefont {Chin}\ and\ \citenamefont {Chen}(2002)}]{Chin2002}%
  \BibitemOpen
  \bibfield  {author} {\bibinfo {author} {\bibfnamefont {S.~A.}\ \bibnamefont {Chin}}\ and\ \bibinfo {author} {\bibfnamefont {C.~R.}\ \bibnamefont {Chen}},\ }\bibfield  {title} {\bibinfo {title} {{G}radient symplectic algorithms for solving the {S}chrödinger equation with time-dependent potentials},\ }\href {https://doi.org/10.1063/1.1485725} {\bibfield  {journal} {\bibinfo  {journal} {The Journal of Chemical Physics}\ }\textbf {\bibinfo {volume} {117}},\ \bibinfo {pages} {1409} (\bibinfo {year} {2002})}\BibitemShut {NoStop}%
\bibitem [{\citenamefont {Sakkos}\ \emph {et~al.}(2009)\citenamefont {Sakkos}, \citenamefont {Casulleras},\ and\ \citenamefont {Boronat}}]{Sakkos2009}%
  \BibitemOpen
  \bibfield  {author} {\bibinfo {author} {\bibfnamefont {K.}~\bibnamefont {Sakkos}}, \bibinfo {author} {\bibfnamefont {J.}~\bibnamefont {Casulleras}},\ and\ \bibinfo {author} {\bibfnamefont {J.}~\bibnamefont {Boronat}},\ }\bibfield  {title} {\bibinfo {title} {{H}igh order {C}hin actions in {P}ath {I}ntegral {M}onte {C}arlo},\ }\href {https://doi.org/10.1063/1.3143522} {\bibfield  {journal} {\bibinfo  {journal} {The Journal of Chemical Physics}\ }\textbf {\bibinfo {volume} {130}},\ \bibinfo {pages} {204109} (\bibinfo {year} {2009})}\BibitemShut {NoStop}%
\bibitem [{\citenamefont {Boninsegni}\ \emph {et~al.}(2006)\citenamefont {Boninsegni}, \citenamefont {Prokof'ev},\ and\ \citenamefont {Svistunov}}]{Boninsegni2006}%
  \BibitemOpen
  \bibfield  {author} {\bibinfo {author} {\bibfnamefont {M.}~\bibnamefont {Boninsegni}}, \bibinfo {author} {\bibfnamefont {N.~V.}\ \bibnamefont {Prokof'ev}},\ and\ \bibinfo {author} {\bibfnamefont {B.~V.}\ \bibnamefont {Svistunov}},\ }\bibfield  {title} {\bibinfo {title} {{W}orm algorithm and diagrammatic {M}onte {C}arlo: {A} new approach to continuous-space path integral {M}onte {C}arlo simulations},\ }\href {https://doi.org/10.1103/PhysRevE.74.036701} {\bibfield  {journal} {\bibinfo  {journal} {Phys. Rev. E}\ }\textbf {\bibinfo {volume} {74}},\ \bibinfo {pages} {036701} (\bibinfo {year} {2006})}\BibitemShut {NoStop}%
\bibitem [{\citenamefont {Pitaevskii}\ and\ \citenamefont {Stringari}(2016)}]{Pitaevskii2016}%
  \BibitemOpen
  \bibfield  {author} {\bibinfo {author} {\bibfnamefont {L.}~\bibnamefont {Pitaevskii}}\ and\ \bibinfo {author} {\bibfnamefont {S.}~\bibnamefont {Stringari}},\ }\href {https://doi.org/10.1093/acprof:oso/9780198758884.001.0001} {\emph {\bibinfo {title} {{{B}ose-{E}instein {C}ondensation and {S}uperfluidity}}}}\ (\bibinfo  {publisher} {Oxford University Press},\ \bibinfo {year} {2016})\BibitemShut {NoStop}%
\bibitem [{\citenamefont {Cikojević}\ \emph {et~al.}(2018)\citenamefont {Cikojević}, \citenamefont {Markić},\ and\ \citenamefont {Boronat}}]{Cikojevic2018}%
  \BibitemOpen
  \bibfield  {author} {\bibinfo {author} {\bibfnamefont {V.}~\bibnamefont {Cikojević}}, \bibinfo {author} {\bibfnamefont {L.~V.}\ \bibnamefont {Markić}},\ and\ \bibinfo {author} {\bibfnamefont {J.}~\bibnamefont {Boronat}},\ }\bibfield  {title} {\bibinfo {title} {{H}armonically trapped {B}ose–{B}ose mixtures: a quantum {M}onte {C}arlo study},\ }\href {https://doi.org/10.1088/1367-2630/aad6cc} {\bibfield  {journal} {\bibinfo  {journal} {New Journal of Physics}\ }\textbf {\bibinfo {volume} {20}},\ \bibinfo {pages} {085002} (\bibinfo {year} {2018})}\BibitemShut {NoStop}%
\bibitem [{\citenamefont {Krotscheck}\ \emph {et~al.}(1988)\citenamefont {Krotscheck}, \citenamefont {Saarela},\ and\ \citenamefont {Epstein}}]{Krot1}%
  \BibitemOpen
  \bibfield  {author} {\bibinfo {author} {\bibfnamefont {E.}~\bibnamefont {Krotscheck}}, \bibinfo {author} {\bibfnamefont {M.}~\bibnamefont {Saarela}},\ and\ \bibinfo {author} {\bibfnamefont {J.~L.}\ \bibnamefont {Epstein}},\ }\bibfield  {title} {\bibinfo {title} {Impurity states in liquid-helium films},\ }\href {https://doi.org/10.1103/PhysRevB.38.111} {\bibfield  {journal} {\bibinfo  {journal} {Phys. Rev. B}\ }\textbf {\bibinfo {volume} {38}},\ \bibinfo {pages} {111} (\bibinfo {year} {1988})}\BibitemShut {NoStop}%
\bibitem [{\citenamefont {Clements}\ \emph {et~al.}(1997)\citenamefont {Clements}, \citenamefont {Krotscheck},\ and\ \citenamefont {Saarela}}]{Krot2}%
  \BibitemOpen
  \bibfield  {author} {\bibinfo {author} {\bibfnamefont {B.~E.}\ \bibnamefont {Clements}}, \bibinfo {author} {\bibfnamefont {E.}~\bibnamefont {Krotscheck}},\ and\ \bibinfo {author} {\bibfnamefont {M.}~\bibnamefont {Saarela}},\ }\bibfield  {title} {\bibinfo {title} {Impurity dynamics in boson quantum films},\ }\href {https://doi.org/10.1103/PhysRevB.55.5959} {\bibfield  {journal} {\bibinfo  {journal} {Phys. Rev. B}\ }\textbf {\bibinfo {volume} {55}},\ \bibinfo {pages} {5959} (\bibinfo {year} {1997})}\BibitemShut {NoStop}%
\bibitem [{\citenamefont {Andreev}(1966)}]{Andreev1969}%
  \BibitemOpen
  \bibfield  {author} {\bibinfo {author} {\bibfnamefont {A.~F.}\ \bibnamefont {Andreev}},\ }\bibfield  {title} {\bibinfo {title} {Surface tension of weak {H}elium isotope solutions},\ }\href {http://jetp.ras.ru/cgi-bin/dn/e_023_05_0939.pdf} {\bibfield  {journal} {\bibinfo  {journal} {Soviet Physics JETP}\ }\textbf {\bibinfo {volume} {23}},\ \bibinfo {pages} {939} (\bibinfo {year} {1966})}\BibitemShut {NoStop}%
\bibitem [{\citenamefont {Lekner}(1970)}]{Lekner1970}%
  \BibitemOpen
  \bibfield  {author} {\bibinfo {author} {\bibfnamefont {J.}~\bibnamefont {Lekner}},\ }\bibfield  {title} {\bibinfo {title} {Theory of surface states of \ce{^{3}He} atoms in liquid \ce{^{4}He}},\ }\href {https://doi.org/10.1080/14786437008220937} {\bibfield  {journal} {\bibinfo  {journal} {The Philosophical Magazine: A Journal of Theoretical Experimental and Applied Physics}\ }\textbf {\bibinfo {volume} {22}},\ \bibinfo {pages} {669} (\bibinfo {year} {1970})}\BibitemShut {NoStop}%
\bibitem [{\citenamefont {Pascual}\ \emph {et~al.}(2024)\citenamefont {Pascual}, \citenamefont {Wasak}, \citenamefont {Negretti}, \citenamefont {Astrakharchik},\ and\ \citenamefont {Boronat}}]{SRFRWA_2024}%
  \BibitemOpen
  \bibfield  {author} {\bibinfo {author} {\bibfnamefont {G.}~\bibnamefont {Pascual}}, \bibinfo {author} {\bibfnamefont {T.}~\bibnamefont {Wasak}}, \bibinfo {author} {\bibfnamefont {A.}~\bibnamefont {Negretti}}, \bibinfo {author} {\bibfnamefont {G.~E.}\ \bibnamefont {Astrakharchik}},\ and\ \bibinfo {author} {\bibfnamefont {J.}~\bibnamefont {Boronat}},\ }\href {https://doi.org/10.18150/SRFRWA} {\bibinfo {title} {{Datasets for ``Temperature-induced miscibility of impurities in trapped Bose gases''}}} (\bibinfo {year} {2024})\BibitemShut {NoStop}%
\end{thebibliography}%

\end{document}